# Planar lightwave circuits for quantum cryptographic systems

Yoshihiro NAMBU, Takaaki HATANAKA, and Kazuo NAKAMURA

We propose a quantum cryptographic system based on a planar lightwave circuit (PLC) and report on optical interference experiments using PLC-based unbalanced Mach-Zehnder interferometers (MZIs). The interferometers exhibited high-visibility (>0.98) interference even when the polarisation in the optical fibre connecting the two MZIs was randomly modulated. The results demonstrate that a PLC-based setup is suitable for achieving a polarisation-insensitive phase-coding cryptographic system.

*Introduction*: Quantum key distribution (QKD) allows two remote parties, say Alice and Bob, to generate a secret key, with privacy guaranteed by quantum mechanics [1-2]. Since Bennett and Brassard's invention of the BB84 protocol [1] and its first demonstration [2], numerous QKD systems have been developed that use optical fibre-based techniques and faint laser pulses [3-10]. In these systems, the relative phase of the superposed single photon wave packets was used to encode the random bit data. Townsend et al. first demonstrated phase coding using unbalanced Mach-Zehnder interferometers (MZIs) and 10-km transmission of single interfering photons [3]. Further progress was made by Muller et al. [8], who used a single MZI and Faraday mirrors to self-align the polarisation and to self-balance the path length of an interferometer, in what they called a plug and play (P&P) system. Kosaka et al. recently demonstrated single photon interference over 100 km by using the P&P system [11].

Although the P&P system works well for QKD systems that use a faint laser pulse as long as the transmission length is less than 100 km, transmission over greater distances is difficult even if a photon detector with extremely high SN characteristics is used [11]. This is because backscattering noise in the fibre dominates the intrinsic noise of the photon detector. The bi-directional nature of the P&P system is responsible for the high backscattering noise [10]. Although the use of a storage line and burst photon trains would reduce the backscattering, it would also reduce the effective transmission rate [10]. A better approach to solving this problem is to use a uni-directional system [3]. A uni-directional system is also desirable to realize more secure QKD system using a true single photon, because one cannot use a brighter pulse from Bob to Alice, as in a P&P system, and the round-trip transmission of a single photon limits the allowable distance between two parties in bi-directional setups to half that in uni-directional setups.

It is, however, not easy, using optical fibre-based techniques, to implement a uni-directional system in which two identical and independently unbalanced MZIs remain stable on the sub-wavelength scale. To solve this problem, a uni-directional system based on a planar lightwave circuit (PLC) has been proposed [12]. In this letter we propose a QKD system designed for PLC platforms and report the results of experiments using PLC-based interferometers.

*PLC-based QKD system:* Our system consists of a light source, a Mach-Zehnder switch (MZSW), and two unbalanced MZIs, one for Alice and one for Bob, connected in series by an optical fibre (Fig.1). The MZIs are composed of PLCs, which provide good stability for the system [12]. The MZSW includes a phase modulator, and is made of $LiNbO_3$ waveguides which can be combined with PLC. By the combined action of the MZSW and the MZI, Alice can randomly prepare one of four states $\{|l\rangle, |s\rangle, |l\rangle+|s\rangle, |l\rangle-|s\rangle\}$ on the optical fibre link, where $|l\rangle$ and $|s\rangle$ denote a single photon state having traveled via the short or the long path of Alice's MZI. After the photons travel through Bob's MZI, Bob finds them in one of three time slots. The first slot corresponds to photons taking the short path in both MZIs, while the last slot corresponds to photons taking both the long paths. The central one corresponds to photons taking the short path in Alice's MZI and the long one in Bob's, and vice versa. If Alice chooses states $|l\rangle$ and $|s\rangle$, her choice will correlate with photon detection in the last and first time slots, respectively. If she chooses the superposed states $|l\rangle \pm |s\rangle$, the two possible paths of the photons in the central slot interfere, and her choice will correlate with the output ports of the photons at Bob's MZI in the central slot. These basis-dependent ($\{|l\rangle, |s\rangle\}$ or $\{|l\rangle+|s\rangle, |l\rangle-|s\rangle\}$) correlations between the state prepared by Alice and the outcome observed by Bob can be used to implement a BB84 QKD system. The feature of our system is that it requires a single random number generator and a single phase-modulator at Alice's site only, not at Bob's. This will increase security and reduce optical loss in Bob's equipment.

*Experiment:* We fabricated PLC-based MZIs made of silica waveguides on silicon substrates. The path-length difference was ~1.6 m, corresponding to a temporal delay of ~8 ns, and the device size was ~100×35 mm. Polarisation-maintaining fibre pigtails aligned to the substrate surface were connected to the input and output of the MZIs. A Peltier element was attached to the back of the substrate to control the device temperature. Optical loss in the MZIs was determined mainly by the loss in the long path of the waveguide and was ~8 dB (excluding the 3-dB intrinsic loss at the coupler). By using current technology, however, the loss will be reduced to less than 3 dB. To achieve a passively stabilized system, the imbalance between the two MZIs must be kept stable to maintain correct phase relations. Furthermore, making the system insensitive to the polarisation rotation that occurs in the optical links allows us to avoid active polarisation control. Polarisation transformation in both paths in Bob's MZI must be kept balanced to achieve this insensitivity. To investigate whether the imbalance was stabilized and the system was insensitive to the polarisation rotation, we observed the interference that appeared in the photons at the two output ports of the second MZI and detected in the central time slot.

The two MZIs were connected using an optical fibre (Fig. 2). A polarisation randomizer was used to simulate random polarisation rotation in the optical link. A TE-polarized faint laser pulse ($\lambda$ of 1.55 $\mu$m; length of ~500 ps) was injected into the first MZI. The repetition rate was 500 kHz, and the photons at the two output ports of the second MZI were detected by two InGaAs avalanche photodiodes (APDs) operating in gated (2.5 ns) photon counting mode. The average number of photons per pulse was set low enough to ensure the linearity of the APDs. After the first MZI, two optical pulses with identical, though randomly changing, polarisation propagated through the fibre link. Temperature T1 of the first MZI was controlled so as to balance the path length difference between the two MZIs, and T2 of the second MZI was adjusted and fixed in a manner such that the arbitrary polarisation state of the incoming pulses undergoes the same transformation in both paths of the second MZI.

Figure 2 shows the measured photon counts during 10-s intervals at the two output ports of the second MZI as a function of T1. Controlling the temperature of the two MZIs enabled us to observe a clear interference fringe with visibility of over 98%, even when the polarization in the fibre link was randomly modulated. The system remained passively stable for over an hour. We also evaluated the system with a 10-km optical fibre link and obtained the same results.

*Conclusion:* We have proposed a QKD system designed for PLC platforms and demonstrated high-visibility interference using two PLC-based unbalanced MZIs even when polarisation in the optical fibre link was randomly modulated. This indicates that the PLC-based setup is suitable for implementing a polarisation-insensitive, passively stabilized phase-coding QKD system.

This work was supported by the Telecommunications Advancement Organization of Japan.

**References**

[1] Bennett, C.H. and Brassard, G.: 'Quantum cryptography: Public key distribution and coin tossing'. Proc. Int. Conf. Comput. Syst. Signal Process., Bangalore, 1984, pp. 175-179.

[2] Bennett, C.H., Bessette, F., Brassard, G., Salvail, L., and Smolin, J.: 'Experimental quantum cryptography', J. Cryptol., 1992, 5, (3).

[3] Townsend, P.D., Rarity, J.G., and Tapster, P.R., 'Single photon interference in 10 km long optical fibre interferometer', Electronics Letters, 1993, 29, pp. 634-635.

[4] Townsend, P.D., Rarity, J.G., and Tapster, P.R., 'Enhanced single photon fringe

visibility in a 10 km-long prototype quantum cryptography channel', Electronics Letters, 1993, 29, pp. 1291-1293.

[5] Franson, J.D. and Jacobs, B.C., 'Operational system for quantum cryptography', Electronics Letters, 1995, 31, pp. 232-234.

[6] Christophe Marand and Paul D. Townsend: 'Quantum key distribution over distances as long as 30 km', Optics Letters, 1995, 20, pp. 1695-1697.

[7] Muller, A., Zbinden, H., and Gisin, N.: 'Quantum cryptography over 23 km in installed under-lake telecom fibre', Europhysics Letters, 1996, 33, pp. 335-339.

[8] Muller, A., Herzog, T., Huttner, B., Tittel, W., Zbinden, H., and Gisin, N.: '"Plug and play" systems for quantum cryptography', Applied Physics Letters, 1997, 70, pp. 793-795.

[9] Zbinden, H., Gautier, J.D., Gisin, N., Huttner, B., Muller, A., and Tittel, W.: 'Interferometry with Faraday mirrors for quantum cryptography', Electronics Letters, 1997, 33, pp. 586-588.

[10] Ribordy, G., Gautier, J.-D., Gisin, N., Guinnard, O., and Zbinden, H. 'Automated `plug and play' quantum key distribution', Electronics Letters, 1998, 34, pp. 2116-2117.

[11] Kosaka, H., Tomita, A., Nambu, Y., Kimura, N., and Nakamura, K.: 'Single-photon interference experiment over 100km for quantum cryptography system using a balanced gated-mode photon detector', quant-ph/0306066, 2003, submitted to Electronics Letters.

[12] Bonfrate, G., Harlow, M., Ford, C., Maxwell, G., and Townsend, P.D., 'Asymmetric Mach-Zehnder germano-silicate channel waveguide interferometers for quantum cryptography systems', Electronics Letters, 2001, 37, pp. 846-847.

**Authors' affiliations:**

Yoshihiro Nambu and Kazuo Nakamura (Fundamental Research Laboratories, NEC, 34 Miyukigaoka, Tsukuba, Ibaraki 305-8501, Japan)

Takaaki Hatanaka (Fiber Optic Devices Division, NEC Corp., 747, Magi, Ohtsuki-machi, Ohtsuki, Yamanashi 401-0016, Japan)

**Figure Captions**

Fig. 1. PLC-based QKD system. LS: light source, MZSW: Mach-Zehnder switch, MZI: unbalanced Mach-Zehnder interferometer, ATT: attenuator, PM: phase modulator, APD: avalanche photodiode.

Fig. 2. Experimental set-up using two PLC-based MZIs and photon count during 10-s intervals at two ports of second MZI against T1.

Figure 1

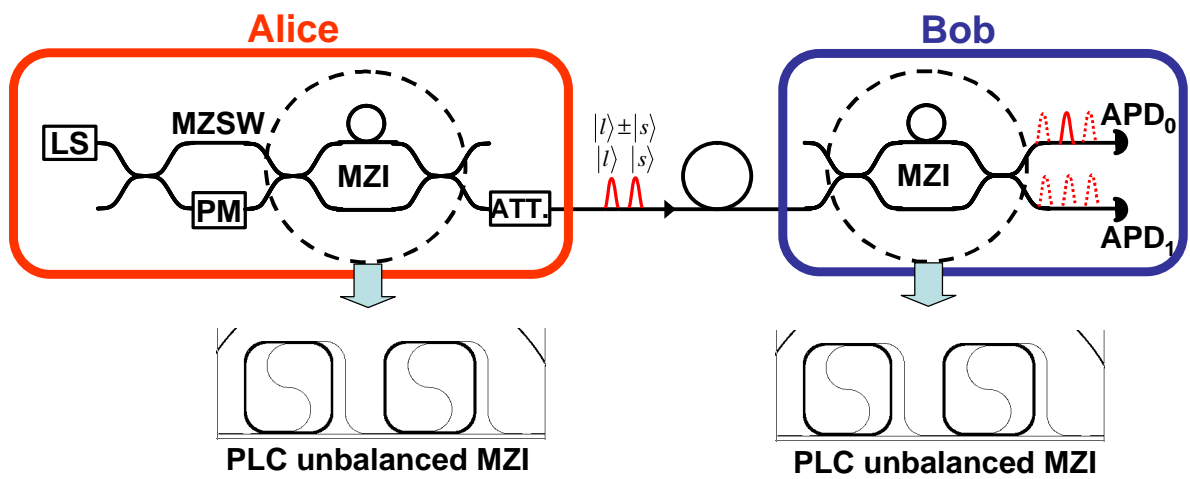

Figure 2

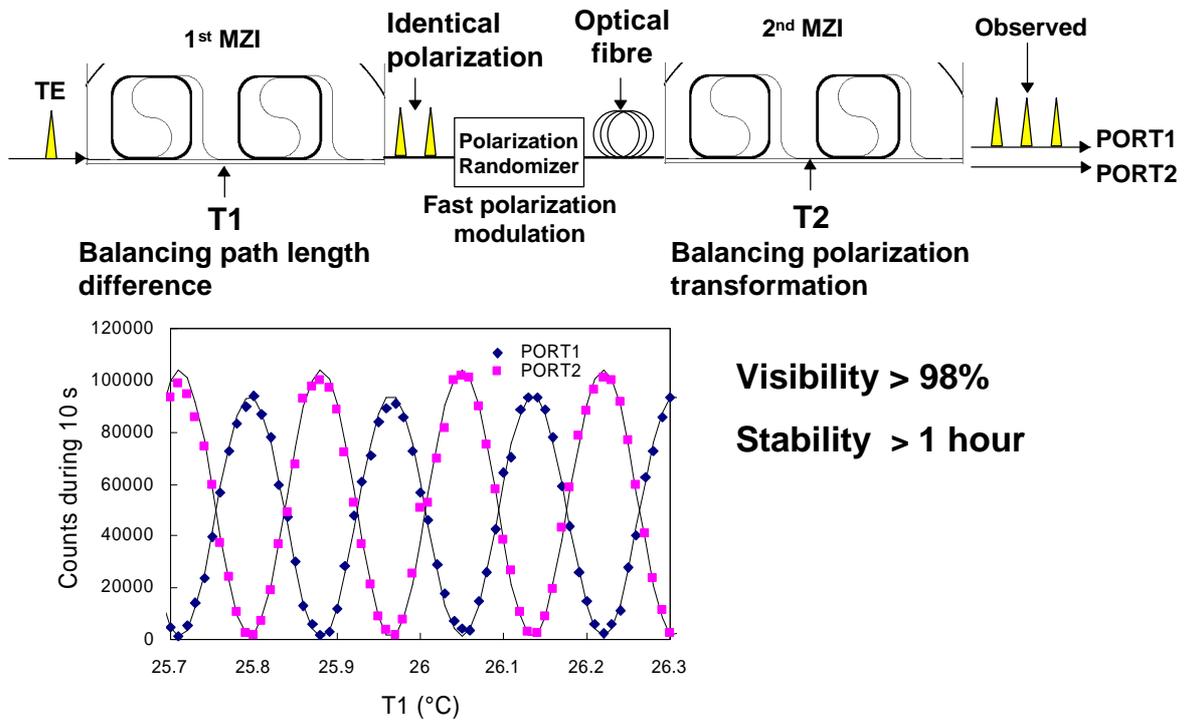